\documentstyle[12pt]{article}
\begin{document}
\hoffset=-1.5 true cm
\voffset=-2 true cm
\setlength{\textheight}{22.0cm}
\textwidth20cm
\title{Sonoluminescence: The Superradiance Paradigm}
\author{M. Schiffer \thanks{e.mail: marcelos@huji.ac.il}\hspace{.2cm}\thanks
{permanent address: IMECC, Universidade Estadual de Campinas, Brazil}
\\Racah Institute, Hebrew University of Jerusalem,\\Givat Ram,
91904, Jerusalem}
\maketitle
\begin{center} {\bf Abstract}
\end{center}
\newcommand{\dv}{\vec{\nabla}\cdot}
\newcommand{\rt}{\vec{\nabla}\times}
\newcommand{\B}{\vec{B}}
\newcommand{\h}{\vec{H}}
\newcommand{\D}{\vec{D}}
\newcommand{\E}{\vec{E}}
\newcommand{\n}{\vec{n}}
\newcommand{\K}{\vec{k}}
\newcommand{\dt}[1]{\frac{\partial{#1}}{\partial t}}
\newcommand{\be}{\begin{equation}}
\newcommand{\ee}{\end{equation}} {\footnotesize In sonoluminescence sound waves 
in the fluid 
are converted into UV-photons. In this letter we put forward a mechanism based 
on superradiance
that operates this conversion. The predicted  energies of emitted light turns 
out 
to lie in the range of 1-10 eV, in good  agreement with experiment. Furthermore 
our paradigm 
hints why the effect  hinges so crucially on the cavitation of a noble gas}
\vspace{1.0cm}

\section{Introduction}

Sonoluminescence -- the effect whereby a gas cavitating in water  whose bubble's 
size is
driven by ultrasound waves glows -- has been puzzling the mind of scientists 
since the
thirties (for a recent review see \cite{review} and for a general account 
\cite{sciam}). 
The bubble starts expanding from an ambient radius of  $5 \mu m$ to $50 \mu m$ 
and then collapses
 at about about 4 times the speed of sound to $0.5 \mu m$. Then, within $50ps$ 
of the  minimum 
radius a burst of light containing about 1 million photons with energies ranging 
from $6.5 eV$ down 
to $2.0eV$ is emitted. So far, there is not a  satisfactory understanding on the 
mechanism involved
in the conversion of phonons in the liquid into UV-photons \cite{review}. The 
most viable 
candidate is the shock wave model \cite{review}, whereby at late stages of 
the bubble's collapse, its surface moves supersonically, producing a shock 
\cite{jar}-\cite{moss}.
In this scenario, the fast collapse heats the gas in the interior of the bubble 
to temperatures of 
the    order $T \sim 10^5$. The gas is ionized and  the hot ionized plasma emits 
via Bremstrahlung.
For $M=v/c \sim 4$ the emitted photons lie in the experimental range. Less 
viable models
involve the emission of light through the electronic discharge owing to a charge 
separation
inside the plasma \cite{Lepoint}, Unruh-Davies radiation from the surface 
\cite{eberlein,knight}, 
and the bursting of light that follows the sudden cracking of a putative ice 
produced by the high
 pressures present inside the bubble \cite{hickling}. In the Unruh-Davies like 
calculation too few
 photons are produced. This difficulty seems to be circumvented by a recent 
Casimir energy 
inspired mechanism, based on the modification of the electromagnetic action 
\cite{chodos}. It is 
essential  to remark that no sonoluminescence is produced unless the gas inside 
the bubble is 
doped with a noble gas \cite{review}. Unfortunately, none of the proposed 
mechanisms can  address 
this fact. 

Superradiance is the amplification of radiation by rotating absorbing media. 
Zel'dovich was
the first to consider this effect in the context of an electromagnetic field 
bouncing off an
ohmic cylinder rotating with angular velocity $\Omega$ \cite{zel1} and also of a 
scalar
field \cite{zel2}. Misner considered a similar effect for quanta scattered by a 

 rotating black hole \cite{misner} . In both instances the condition $\omega - m 
\Omega < 0$ 
must be satisfied, where $\omega$ and $m$ are the frequency and azimuthal 
angular momentum of 
the emitted quanta.
It was already known by Ginzburg and Frank \cite{ginz} that superradiance and 
Cherenkov emission 
are interchangeable concepts and  rely upon the fact that the modulus of the 
velocity $\vec{v}$
of the source of radiation exceeds the phase velocity of  wave in the medium, in 
which case 
the frequency and wave number of the wave satisfy $\omega - \vec{k} \cdot 
\vec{v} < 0$. Apart
from black holes, there are a myriad of instances of superradiant emission: the 
onset of
viscous flow inside  superfluid  through the excitation of phonons, transition 
radiation, to
name just a few \cite{beksch}.  A particularly enlightening example for our 
purposes is the
spontaneous emission of sound waves by a homogeneous medium  that has a 
tangential
discontinuity in the velocity field \cite{landau6}. The reflection coefficient 
of a sound wave
incident upon the the surface of discontinuity in the velocity field has poles 
for some values
of the incidence angle, which are physically interpreted as the angles where 
sound waves are
spontaneously emitted. The onset of  emission is governed by the condition 
$M=v/c >1$
(actually $M> 2$, but the tighter bound stems from a  kinematical rather than 
dynamical
constraint). The above scenario could easily explain the emission of  sound 
waves during the
supersonic contraction of the bubble, but how could this possibly relate  to the 
emission of light
which apparently requires superluminal motion? 

In the scenario we put forward, the formation of a shock wave inside the  bubble 
is pivotal.
Once it is formed,  the gas passing through the shock is compressed, that is to 
say, a
density discontinuity is generated across the shock
\cite{landau6} which, in turn,  leads to a discontinuity in the dieletric 
permittivity   (we
assume that the permeability $\mu =1$). In analogy with the previous example, 
the
reflection coefficient for originally incoming waves also display poles at the 
denominator
for  given frequencies. These are responsible for the (classical)  stimulated 
emission of
light. In what follows we  consider a somewhat  simplified model, a plane shock 
moving with
constant velocity. 

\section{The model}

Our first step is to study the propagation of electromagnetic waves in  moving 
media. The
appropriate equations  for  nonrelativistic motion contains a correction of 
order $v/c$ to
Maxwell's equations. Writing the velocity as $\vec{v} = v\n$ where $\n$ is the 
normal vector
pointing, say,  to the right \cite{landau8}:
\begin{eqnarray}
 \dv \B &=& 0 ; \hspace{0.5cm}  \rt \E = - \dt{\B}  \nonumber \\
 \dv \D &=& 0 ;  \hspace{0.5cm}  \rt \h = \dt{\D} \\
\mbox{and}\\
\D &=&\epsilon \E + \alpha \n \times \h \nonumber \\
\B &=& \h - \alpha \n \times \E  
\end{eqnarray}  where $\alpha =(\epsilon -1) v$. Taking a plane wave $\exp 
i(\K\cdot \vec{x} -
\omega t)$ it follows that,
\begin{eqnarray} 
A \E + B \n \times \h &=& 0 , \nonumber\\ 
B \epsilon \E + A \n \times \h &=&0
\end{eqnarray}  where $A=\epsilon \omega^2 - k^2 - \alpha k \omega$ and 
$B=\alpha\epsilon$.
For compatibility $A^2 -\epsilon B^2=0$. Thus, to the leading order 
\be
\E = \mp \frac{1}{\sqrt{\epsilon}} \n \times \h   \hspace{.5cm} \mbox{with} 
\hspace{.5cm}  
\left|k\right| =(\ \sqrt{\epsilon} \pm \alpha) \omega
\label{dispersion}
\ee  
The $\mp /\pm$ signs refer to right and left moving waves , respectively. In the 
picture
below  the shock is moving to the right, corresponding to the collapse of the 
bubble. 
%\begin{picture}\special{draw.eps} \end{figure}
 The above equations must be supplement by the boundary conditions. To the same 
order in
$v/c$ \cite{landau8} we have continuity of the  normal components at the shock
\be
\Delta \D_n = 0 ; \hspace{.5cm} \Delta \B_n = 0 ,
\ee
but 
\be
\n\times \Delta\E = v \Delta \B_t ; \hspace{.5cm} \n\times \Delta\h =-v \Delta 
\D_t 
\ee 
Because $\mu=1$, $\Delta \B_t = \Delta \h_t= {\cal{O}}(v/c)$ and  
$\Delta \D_t = \Delta (\epsilon \E_t)$. Thus, up to first  order in $v/c$ we 
have:
\begin{eqnarray}
\Delta \E_t &=& 0 \nonumber \\ n\times \Delta\h &=& -v \Delta (\epsilon E_t)
\label{dh}
\end{eqnarray} 
Let us now consider a (right-moving) wave propagating from the exterior of the
shock to its interior. Let 
$(\E_0,\h_o),(\E_1,\h_1)$ and $(\E_2,\h_2)$ represent the incoming, reflected 
and transmitted
waves, respectively. Let us further assume that the incidence is normal to the 
shock (the effect 
subsists for non-normal incidence too, but is less intense). Under these  
circumstances  $\E_2
= \E_0+\E_1$, and substituting into eq. (\ref{dh}), it follows that
\be 
n\times (\h_2-\h_1-\h_0) = -v \left((\epsilon_2 -\epsilon_1) \E_0 \, .
+(\epsilon_2-\tilde{\epsilon_1}) \E_1 \right) 
\ee 
In the above equation, due care of redshifts of the reflected and transmitted 
waves was
taken into account in  the evaluation of the electric permeabilities $
\epsilon_1=\epsilon_1(\omega), \tilde{\epsilon_1}= 
\epsilon_1(\tilde{\omega}),  \epsilon_2 = \epsilon(\omega_2)$, where the 
frequencies are connected 
via the relations:
\be
\omega - k v =\tilde{\omega} + \tilde{k} v = \omega_2 - k_2 v .
\ee 
With the aid of the dispersion relation [eq. (\ref{dispersion})],  to the 
leading order
${\cal{O}}(v/c)$, it follows that:
\be 
\tilde{\omega}= \omega \left[1 - 2 v \sqrt{\epsilon_1(\omega)}\right] , 
\hspace{.5cm}
\omega_2 =\omega \left[1- v \left(\sqrt{\epsilon_1(\omega)} -
\sqrt{\epsilon_2(\omega)}\right)\right] \, .
\label{expansion}
\ee 
Further use of eq.(\ref{dispersion}) yields $n\times (\h_2-\h_1-\h_0)=
(-\sqrt{\epsilon_2(\omega_2)} \E_2 - \sqrt{\epsilon_1(\tilde{\omega})}\E_1+
\sqrt{\epsilon_1}\E_0)$ and the following reflection amplitude:
\be
\frac{E_1}{E_0} =\frac{\sqrt{\epsilon_1(\omega)} - \sqrt{\epsilon_2(\omega_2)} + 
v (
\epsilon_2(\omega_2) - \epsilon_1(\omega))} {\sqrt{\epsilon_2(\omega_2)} +
\sqrt{\epsilon_1(\tilde{\omega_1}}) - 
v (\epsilon_2(\omega_2) - \epsilon_1(\tilde{\omega_1}))} \, .
\ee 
Next, with the aid of eq.(\ref{expansion}), we expand this amplitude to first 
order in
$v/c$, which yields
\be
\frac{E_1}{E_0} =\frac{\sqrt{\epsilon_1} - \sqrt{\epsilon_2} + v 
\left(\epsilon_2 - \epsilon_1
+  \frac{\sqrt{\epsilon_1} - \sqrt{\epsilon_2}}{2\sqrt{\epsilon_2}} \omega 
\frac{\partial
\epsilon_2}{\partial \omega}\right)} {\sqrt{\epsilon_2} + \sqrt{\epsilon_1} - v
\left(\epsilon_2  - \epsilon_1 + \frac{\sqrt{\epsilon_1} - 
\sqrt{\epsilon_2}}{(2\sqrt{\epsilon_2}}
\omega \frac{\partial\epsilon_2}{\partial \omega}\ + \omega \frac{\partial
\epsilon_1}{\partial \omega}\right)} \, .
\ee 
Spontaneous emission is associated with the poles of this reflection amplitude. 
With
$\epsilon \sim 1$ and  the velocity of the shock  of the order of the speed of 
sound, $v \sim
10^{-5}$, makes impossible to find a root of the denominator
\be 
f(\omega)= \sqrt{\epsilon_2} + \sqrt{\epsilon_1} - v \left(\epsilon_2  - 
\epsilon_1 +
\frac{\sqrt{\epsilon_1} - 
\sqrt{\epsilon_2}}{(2\sqrt{\epsilon_2}}
\omega \frac{\partial\epsilon_2}{\partial \omega}\ + \omega \frac{\partial
\epsilon_1}{\partial \omega}\right) \, .
\ee 
if it were not for the following very fortunate fact.  As a consequence of 
having all
electronic shells filled up,  noble gases do not show absorption lines for a 
broad frequency
band, that is to say,  are transparent for $\omega_1 << \omega << \omega_2$. It 
follows from the 
Kramers-Kronig relation that in this interval the medium behaves like a plasma 
and
the corresponding expression for the permittivity is \cite{landau8}:
\be
\epsilon(\omega) = a - \left(\frac{\omega_0}{\omega}\right)^2
\ee 
where $\omega_0 = \frac{4 \pi n e^2}{m}$ with $n$ being the electronic density 
and $e, m$
the electron charge and mass, respectively. Consequently, the permittivity 
vanishes in this 
interval for some frequency $\tilde{\omega} = a^{-1/2} \omega_0$ \cite{landau8}. 
Presuming that
 we are dealing with a weak shock wave (small discontinuities in pressure 
$\delta p$, number 
density $\delta n$ and normal velocity $\delta v$) it follows that:
\begin{eqnarray}
\epsilon_2 \approx 2 a (\frac{\omega}{\tilde{\omega}}-1) \\
\epsilon_1 \approx 2 a \frac{\delta n}{n} + 2 a 
(\frac{\omega}{\tilde{\omega}}-1)
\end{eqnarray}
where we omitted terms of order $\frac{\delta n}{n} 
(\frac{\omega}{\tilde{\omega}}-1)$.  
Inserting these approximations into $f(\omega)$ we plotted $f(\omega)$ for a 
wide range of 
values for  $v \sim 10^{-5}$ and $10^{-2} < \frac{\delta n}{n} < 10^{-4}$. 
Notice that for all cases
the root of $f(\omega)$ lies very close to $\tilde{\omega}$. 

Estimating the emission frequency. For a gas at pressures of 1 atm and room 
temperature, 
$n \sim 10^{20}$ electrons$/cm^3$ which gives $\omega_0 \sim 10 eV$, in good 
agreement with the 
experimental value of $6.5eV$ for the
high energy cutoff. The broad band spectrum could then be due to the change in 
velocity of
the bubble, the change in the density during the collapse and also on the 
dependence of the
poles on the incidence angle (for non-normal incidence).  Note that our model 
does not require
 high temperatures inside the bubble $T \sim 10^5$. This is a most welcome 
result because such 
high temperatures would cause the bubble to emit via bremstrahlung for any gas. 
Furthermore,
the presence of a ionized plasma would imply on  cyclotron emission in the 
presence of a 
sufficiently  large magnetic field. For  $B=20 T$ the cyclotron period is $1 
ps$, 
which is much shorter  than the present bound on the flashwidth and no change in 
the emitted
light was observed \cite{review}. 

Unfortunately our model does not  address the following important issue.  
Sonoluminescence was 
originally discovered for air cavitating in water. When trying to reproduce the 
effect, 
through bubbles containing a mixture of $N_2$ and $O_2$  null results were 
obtained. Only after
doping this mixture with $Ar$ it was possible to make the bubble glow. It is 
nowadays known that
the effect does not occur for any mixture of diatomic gases. It does occur for 
noble gases 
bubbles or for a mixture of diatomic gases doped with some noble gas. The 
presence 
of non-noble gases produces absorption of light in the frequency range of 
interest and void our 
argument. So it seems that the present proposal  needs to be supplemented by 
some mechanism that 
keeps the gas surrounding the shock to be transparent in a wide range of 
frequencies. Whether the
 superradiance paradigm  is a viable  alternative to the bremstrahlung model can 
one be answered 
after  our model is intertwined  with the hydrodynamical theory.

\section*{Acknowledgments} I wish to express my thanks to J. D. Bekenstein, M. 
Rockni 
and R. Zamir for enlightening conversations. I am thankful to {\it FAPESP} 
(brazilian agency)
for financial support.

\end{document}